# Effect of Precursor Stoichiometry on the Performance and Stability of MAPbBr$_3$ Photovoltaic Devices

*Lukas M. Falk, Katelyn P. Goetz, Vincent Lami, Qingzhi An, Paul Fassl, Jonas Herkel, Fabian Thome, Alexander D. Taylor, Fabian Paulus and Yana Vaynzof\**

L. Falk, Dr. K. Goetz, V. Lami, Q. An, P. Fassl, J. Herkel, F. Thome, Dr. A. Taylor, Dr. F. Paulus, Prof. Y. Vaynzof

Kirchhoff Institute for Physics, Im Neuenheimer Feld 227, 69120, Heidelberg, Germany
Centre for Advanced Materials, Im Neuenheimer Feld 225, 69120, Heidelberg, Germany
E-mail: vaynzof@uni-heidelberg.de

The wide band gap methylammonium lead bromide perovskite is promising for applications in tandem solar cells and light-emitting diodes. Despite its utility, there is only a limited understanding of its reproducibility and stability. Herein, the dependence of the properties, performance, and shelf storage of thin films and devices on minute changes to the precursor solution stoichiometry is examined in detail. Although photovoltaic cells based on these solution changes exhibit similar initial performance, the shelf-storage depends strongly on the precursor solution stoichiometry. While all devices exhibit some degree of healing, the bromide-deficient films show a remarkable improvement, more than doubling in their photoconversion efficiency. Photoluminescence spectroscopy experiments performed under different atmospheres suggest that this increase is due in part to a trap healing mechanism that occurs upon exposure to the environment. Our results highlight the importance of understanding and manipulating defects in lead halide perovskites to produce long-lasting, stable devices.

1. Introduction

Hybrid organic-inorganic lead halide perovskites have earned a lot of research attention in the past decade due to their broad-spectrum absorption and efficient photocurrent generation in solar cells, with record performances reaching those of silicon (24.2% photoconversion efficiency, PCE, at the time of writing)[1]. In order to reach commercial potential, however, perovskite solar cells must achieve these high values reliably from device to device, and furthermore, must retain their performance at least long enough to recoup production costs. Despite predictions of high defect tolerance[2,3], these two aspects remain elusive, with many factors playing overlapping roles in the device behavior over time. It is known, for example, that the interactions at the device interfaces can contribute to device breakdown[4–7]. Another key aspect determining device stability is related to the microstructure of the perovskite active layer[8], with large, uniform grains having been shown to be more stable than small grains upon exposure to both oxygen[9] and humidity[10]. The composition of the active layer also plays an important role, with multi-cation compositions showing an overall higher stability[11–13]. Even upon taking these factors into account, literature reports concerning device stability vary greatly even for devices fabricated from the same recipe in the same device architecture[14]. This suggests that stability and reproducibility issues share a link, in which variation in the reproducibility of device performance may also lead to variation in its stability. For example, certain recipes for fabrication of perovskite layers may result in non-homogenous films, which may also serve to increase sample-to-sample variation[15]. Recently, our group uncovered one key factor that adversely impacts the reproducibility and stability of MAPbI$_3$ perovskite solar cells: the exact stoichiometric ratio of the precursor solution. We showed that purposefully adjusting the ratio between the precursor components (methylammonium iodide (MAI) and lead acetate trihydrate (Pb(Ac)$_2$)) in small, almost negligible amounts, results in large variations in the subsequent photovoltaic (PV) performance and stability[16]. These stoichiometric variations are correlated with the photoluminescence (PL) behavior of the MAPbI$_3$ thin films, displaying significant differences in both the initial PL quantum efficiency (PLQE) and its evolution after exposure to light and oxygen[17].

To date, much of the work on stability and performance has focused on the high achievers of the perovskite PV family: MAPbI$_3$ and the triple cation film composition. While solar cells using methylammonium lead tribromide (MAPbBr$_3$) perovskites show much lower PCE, its wide band gap and corresponding high $V_{OC}$ offer the potential for inclusion into tandem cells, where a MAPbBr$_3$ absorber is combined with a narrow band gap material in order to collect additional photons and achieve higher performance[18–20]. Furthermore, its light emission at 545 nm is ideal for application in light-emitting diodes (LEDs), as green is one of the fundamental pixel colors[21–25]. Despite the critical importance of reproducibility and stability for MAPbBr$_3$ films and devices, few reports exist addressing these issues. Apart from early works that suggest MAPbBr$_3$ to be stable upon exposure to light and elevated temperature[26,27], no systematic studies addressing the stability and reproducibility of such films could be found.

In this report, we carefully examine the effect of precursor solution stoichiometry on the properties, performance and storage stability of MAPbBr$_3$ thin-films and devices. By deliberately and incrementally adjusting the ratio of MABr to Pb(Ac)$_2$ in the precursor solution, similar to our previous work on MAPbI$_3$, we tune the composition of the film from slightly bromide-deficient to bromide-excessive, and examine the response of the active layer in PVs and LEDs. While the initial performance of devices shows little dependence on the stoichiometry of the precursor solution, the evolution of their performance upon storage varies drastically. Remarkably, the slightly understoichiometric bromide-based films show a large degree of defect healing that is evident in both the PLQE and device performance, resulting in a significant increase in its PCE upon storage. Our results underline the strong role that film composition plays in defining the properties, performance, and stability of devices, and further promote the idea that defect engineering may be a viable strategy to produce desirable properties in perovskites.

## 2. Results and Discussion
### 2.1. Optoelectronic Properties

To tune the composition of the $MAPbBr_3$ films, we employed the same strategy as was previously used for $MAPbI_3$[16,17], based on the one-step lead-acetate method for film fabrication, as depicted schematically in **Figure 1a**[28]. This recipe has been shown to produce compact films with a highly uniform composition[15,29], reducing pixel-to-pixel or spot-to-spot variation in the measurements[30]. Here, the lead acetate and methylammonium bromide precursors are weighed and dissolved in DMF such that the molar ratio of $MABr:Pb(Ac)_2$ (denoted as $y$) is under the "ideal" stoichiometry of 3 – i.e. $y = 2.95$. Following the fabrication of the devices at $y = 2.95$, a small amount of MABr stock solution in DMF is added to the precursor solution such that the stoichiometry is now 2.97 $MABr:Pb(Ac)_2$. By repeating this method, we create a series of films constituting the profile shown in Figure 1a: $y = 2.95, 2.97, 2.99, 3.0, 3.01,$ and $3.03$ $MABr:Pb(Ac)_2$. This range was selected as it represents an error in precursor solution stoichiometry of below 2%, making it small enough to be possibly introduced unintentionally during device fabrication. A more detailed description of this precursor preparation procedure is included in the supplementary information. This progression of solutions was used to create two types of thin-film: one being Glass/ITO/PEDOT:PSS/$MAPbBr_3$ for the fabrication of solar cells and LEDs (complete device in the inverted architecture with top extraction layer and electrode shown in Figure 1a), and the other being Glass/$MAPbBr_3$, used for optical measurements. Together, these allow for a variety of measurements to understand the impact of precursor solution stoichiometry on thin-film composition and properties in $MAPbBr_3$ films.

One possible change that might be observed is the broadening or narrowing of the optical gap, which would be indicative of changes to the electronic structure of perovskite. As can be seen in **Figure 1b**, the UV-Vis absorption of each film, the absorption onset at 550 nm and peak position near 530 nm is independent of solution stoichiometry; therefore, the band gap is tolerant to the range of error in precursor composition introduced by varying *y*. The normalized photoluminescence spectra (**Figure 1c**) agree with this observation, demonstrating the same peak position and shape for each film. This is also the case for MAPbI$_3$, where both under- and over-stoichiometric films show an absorption onset of about 780 nm[16].

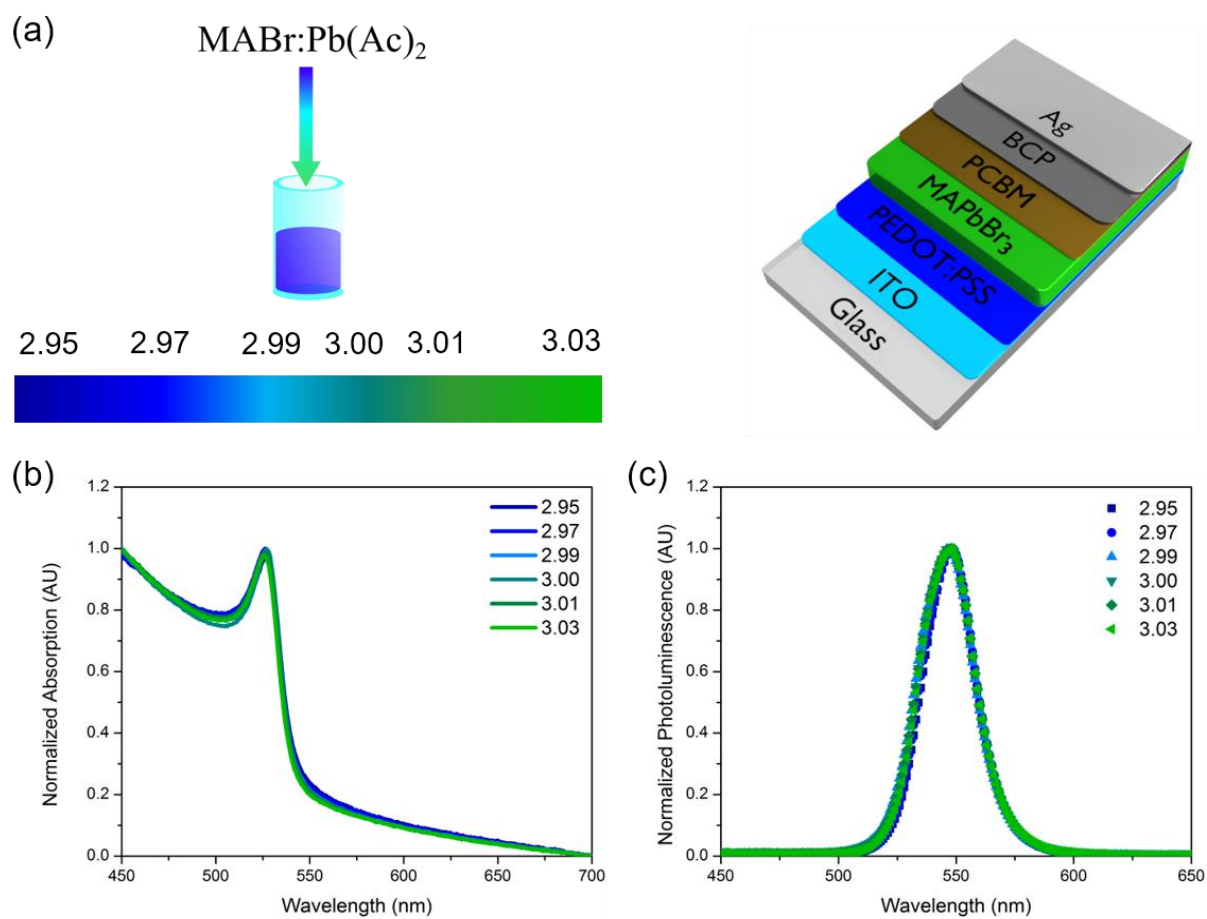

**Figure 1:** (a) Schematic representation of stoichiometry variation and photovoltaic device structure, (b) UV-vis absorption spectra and (c) photoluminescence spectra of MAPbBr$_3$ films with stoichiometry of 2.95, 2.97, 2.99, 3.00, 3.01 and 3.03.

The impact of the precursor solution stoichiometry on film surface composition is, however, directly observed by X-ray photoelectron spectroscopy (XPS), as shown in **Figure 2a**. Here, the intensity of the Pb4f$_{7/2}$, Br3d$_{5/2}$, and N1s peaks was tracked for seven spots on two films of each *y*, allowing us to qualitatively assess the surface uniformity in addition to surface chemistry (with the full survey being shown in the supplementary information, **Figure S1**). As shown by the triangles in Figure 2a, the ratio between the atomic percentages of Br:Pb increases slightly with increasing MABr:Pb(Ac)$_2$ in solution, ranging from 3.75 to a maximum of 4.1. Similarly, the amount of methylammonium at the surface, deduced by tracking the ratio of the N to Pb, also increases, ranging from 1.55 to a maximum of 1.75 for the overstoichiometric films. These trends agree with those observed for MAPbI$_3$; however, the I:Pb and N:Pb ratios in these films increase more rapidly – rising at a rate of 0.075 per 0.01 change in *y*, whereas the bromide films only rise at a rate of 0.05 per 0.01 change in *y*. Comparable error bars indicate that the surface of the films is uniform both across a single sample and between multiple samples. Notably, the ionization potential (**Figure 2b**) is largely unchanged over the range of stoichiometries presented here. This contrasts the results obtained for MAPbI$_3$ films, where the ionization potential increases monotonically by 0.3 eV over the range $\Delta y = 0.12$[16].

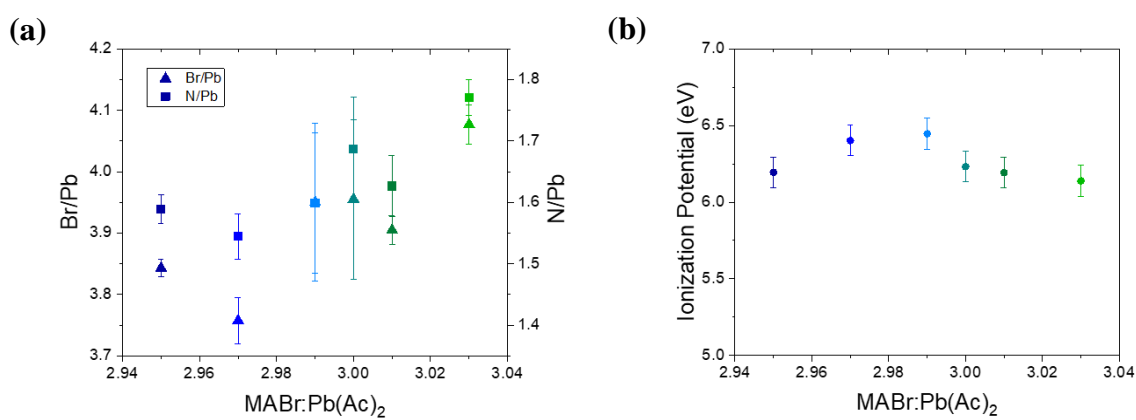

**Figure 2:** (a) Br/Pb (squares) and N/Pb (circles) ratios measured by X-ray photoemission spectroscopy and (b) ionization potential measured by Ultra-Violet photoemission spectroscopy for MAPbBr$_3$ films with different stoichiometry.

## 2.2. Microstructure

The microstructure of films often corresponds to properties observed; for example, a high $J_{SC}$ often correlates with a large grain size[31]. Therefore, we evaluated the surface structure of the films via scanning electron microscopy, shown in **Figure 3**. Unlike its iodide counterparts, where the microstructure is largely unchanged over $\Delta y = 0.1$[16,17], the microstructure of the MAPbBr$_3$ films (where $\Delta y = 0.08$) changes heavily as a function of solution stoichiometry. The understoichiometric ($y = 2.95\text{-}2.99$) films are smooth at the surface and lacking in well-defined grains; but as $y$ increases, grains start to appear, ranging from nm scale to 1 µm in size. The films for all $y$ exhibit complete substrate coverage, lacking in pinholes.

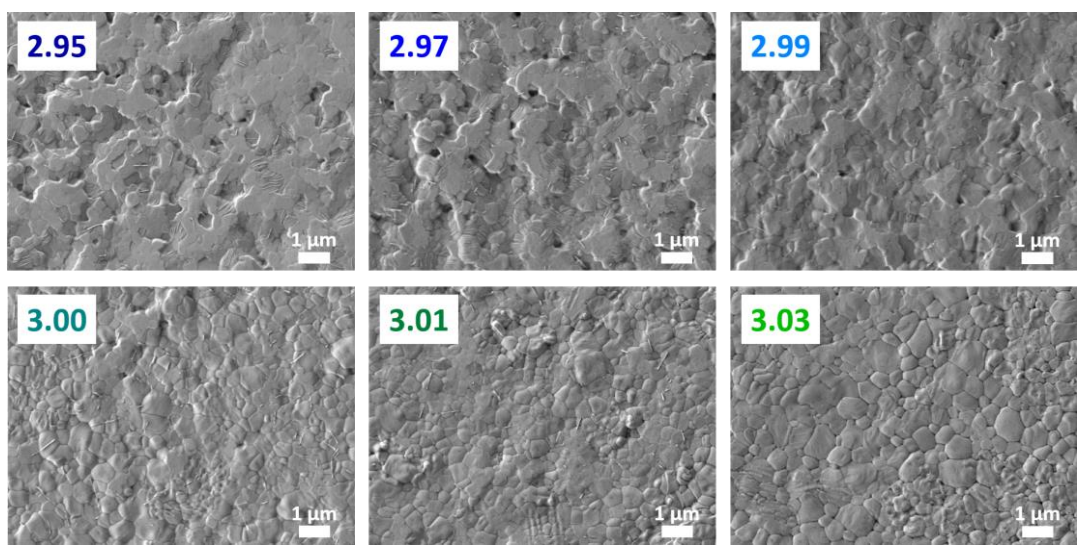

**Figure 3:** Scanning electron microscopy images of MAPbBr$_3$ films with different stoichiometry.

## 2.3. Photovoltaic Performance and Stability

Despite the changes in film microstructure and chemical composition at the surface, the initial performance of photovoltaic cells is highly tolerant to changes in $y$ (**Figure 4,** with device structure shown in Fig. 1a). The open-circuit voltage ($V_{oc}$, Fig. 4b) of the understoichiometric films is somewhat lower than that of the overstoichiometric films; however, this difference approaches the sample-to-sample variation. This observation coincides with the lack of change observed in the ionization potential (Figure 2b) and bandgap (Figure 1b,c). The short-circuit current density ($J_{sc}$), fill-factor (FF), and PCE are all approximately constant as *y* varies. These results contrast the MAPbI$_3$ films, where an increase in the MAI:Pb(Ac)$_2$ ratio (and the I:Pb ratio on the surface of the films) coincides with an increase in the $V_{OC}$, varying linearly over a range of 0.2 V for $\Delta y = 0.1$, with the PCE following suit[16].

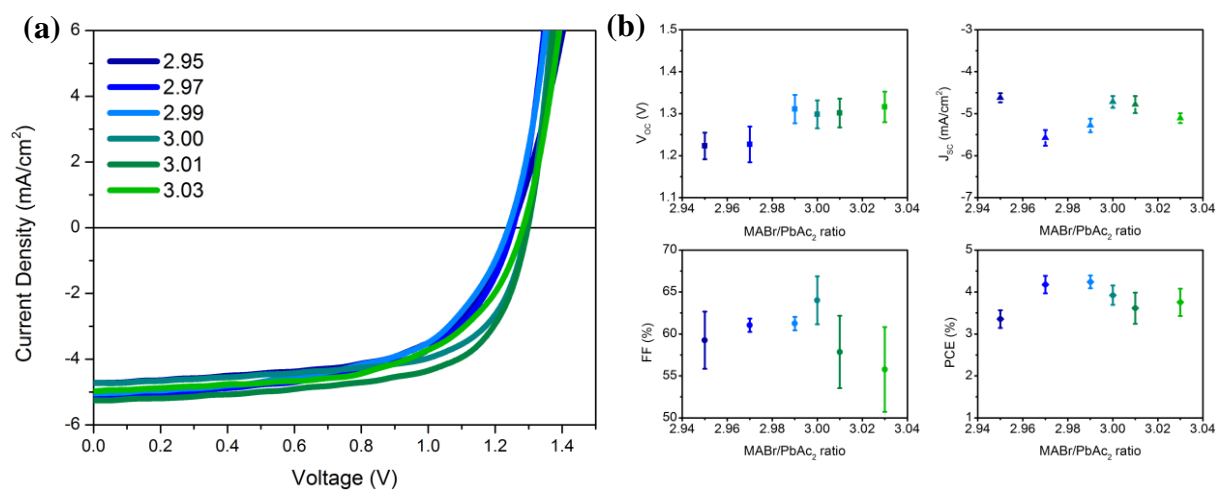

**Figure 4:** (a) J-V characteristics of representative photovoltaic devices with the structure Glass/ITO/PEDOT:PSS/MAPbBr$_3$/PCBM/BCP/Ag. The stoichiometry of the MAPbBr$_3$ active layer was varied between 2.95 and 3.03. (b) Statistics of the photovoltaic performance of the devices.

Following the initial measurements, the devices were covered and placed on a shelf in the lab, to be remeasured again at 10- and 24-day intervals. Interestingly, they display stark differences due to *y* when aged over the course of several weeks. As shown in **Figure 5a**, the change in $V_{OC}$ over time depends on the precise film composition, with the $y = 2.95$ films continually increasing over the time period measured. The $V_{OC}$ for other *y* diminishes over time, or diminishes and then heals slightly, with no observable trend. The fill factor remains roughly constant within the pixel-to-pixel variation. The most apparent trend appears when examining the $J_{sc}$. For higher *y*, the current improves slightly – by about 1 mAcm$^{-2}$. As *y* decreases, such that the precursor solution and films contain less Br, the $J_{sc}$ improvement is much more drastic.

The most understoichiometric film, $y = 2.95$, has an initial value of -4.5 mAcm$^{-2}$; at 24 days, this value has nearly doubled at -8.5 mAcm$^{-2}$. These strong changes result in the clear trend observed in the PCE, where the initial performance for all $y$ is between 3.5 and 4%, but at 10 and 24 days, the $y = 2.95$ films have more than doubled, showing a PCE of almost 8%. The highest stoichiometry films ($y = 3.03$) only increase from 4% to 5% PCE over this time period, with the change in PCE versus $y$ displaying a linear increase with decreasing Br content. For longer shelf storage of over 100 days, the films retain this property of highly increased performance for low $y$ and slightly increased performance for high $y$, suggesting the possibility for long-lived devices when combined with proper encapsulation strategies. This shelf-storage behavior is very different for MAPbI$_3$ films, where the V$_{oc}$ increases between 10 and 20 days for all but the highest stoichiometry studied ($y = 3.075$), and the J$_{sc}$ and PCE both decrease over time for all $y$ (i.e. no such healing is observed)[16].

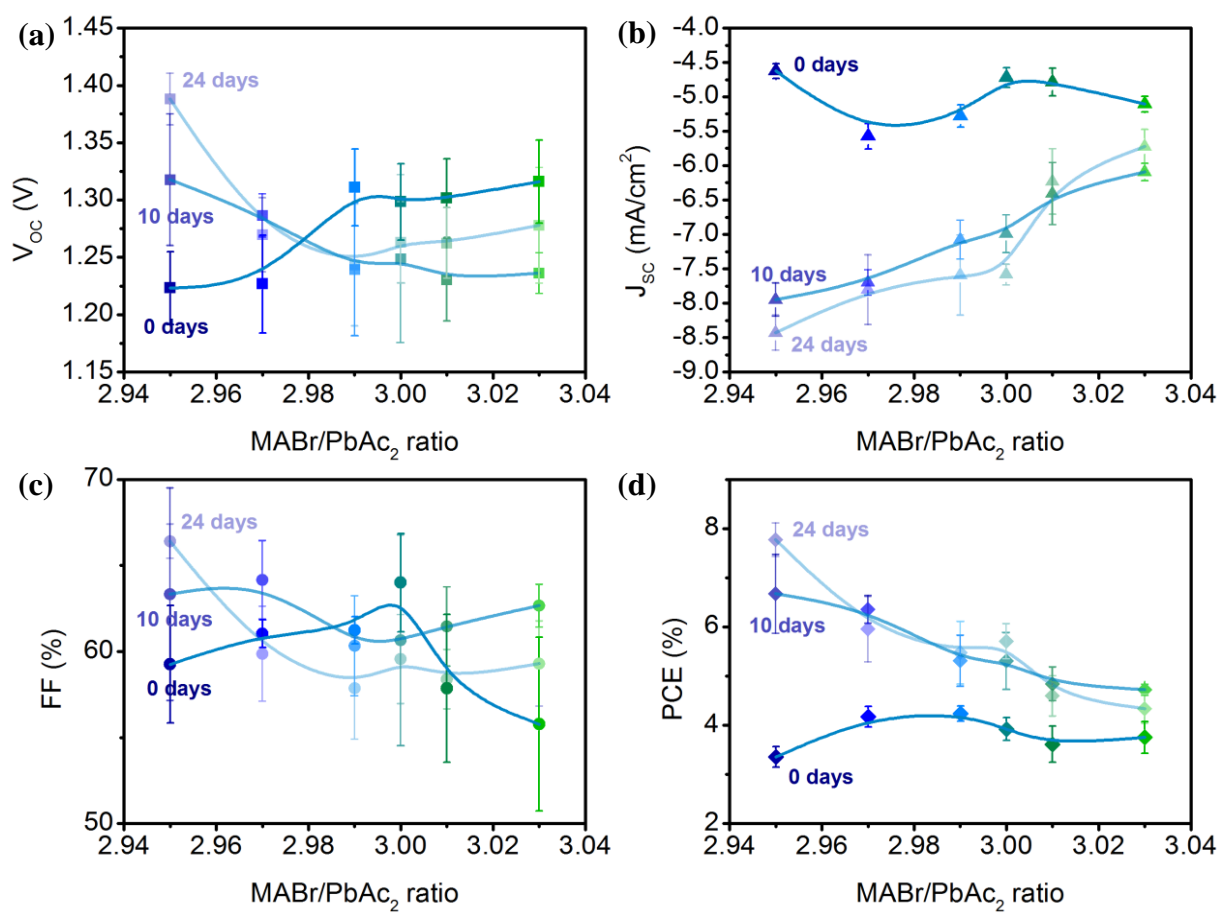

**Figure 5:** Evolution of the photovoltaic performance after 10 and 24 days of shelf storage for devices with different stoichiometry: (a) open-circuit voltage, (b) short-circuit current density, (c) fill factor and (d) power conversion efficiency.

When operated as LEDs, the MAPbBr$_3$ devices show similar trends in their maximum electroluminescent quantum efficiency (ELQE) (**Figure 6a**). Initial measurements are constant at 0.01%; but after 24 days, the $y = 2.95$ films improve to 0.11 %, while the $y = 3.03$ films only improve slightly, to 0.02 %. To gain some insight into what mechanisms might be causing this healing, we measured the PLQE under 405 nm continuous excitation with a power density of ~80 mWcm$^{-2}$ over time and under different atmospheres. As can be seen in **Figure 6b**, the understoichiometric films display the highest PLQE, with $y = 2.95$ at 10 %. The PLQE decreases with increasing $y$, with $y = 3.03$ exhibiting 3% PLQE. These values are consistent with observations of the iodine films, where the understoichiometric films exhibited higher PLQE than the overstoichiometric films[16]. Such behavior is likely a result of the increased formation probability of deep trap states for halide-rich films[2,32], increasing the rate of non-radiative recombination. For the bromide films, the PLQE remains constant through 20 minutes of continuous measurement under nitrogen flushing. When the atmosphere is switched to dry air, the understoichiometric films show rapid improvement, while the overstoichiometric films show only very small increases in PLQE or no change at all. We note that for all stoichiometries we observe no shift in the PL peak position or change in its spectral shape throughout the experiment (**Figure S2**). This suggests that the changes in the PLQE are associated with defect healing, rather than changes in the emission properties of the perovskite layers.

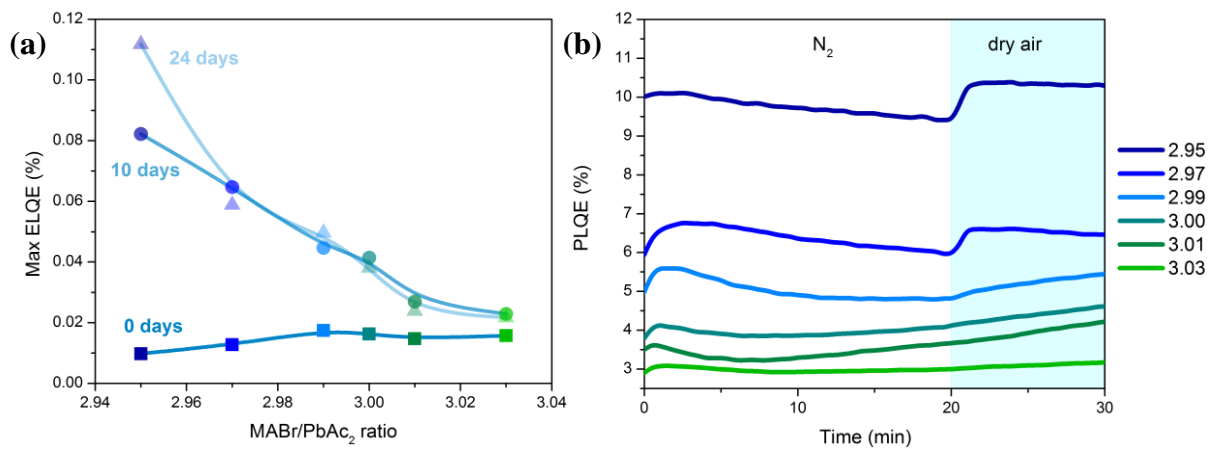

**Figure 6:** (a) Evolution of maximum measured ELQE of the photovoltaic devices presented in Figure 5. (b) Evolution of the PLQE of MAPbBr$_3$ films during exposure to N$_2$ (first 20 min) followed by exposure to dry air under continuous illumination.

This PLQE healing behavior is also seen for MAPbI$_3$ films under exposure to oxygen atmosphere[16,17]. A possible explanation has been attributed to the diffusion of oxygen into iodine vacancy sites, and the subsequent formation of a superoxide species[33,34]. This species is noted to be the appropriate size to fill the vacancy and is thought to be responsible for an initial boost in PLQE under oxygen exposure (and the later decomposition of the film into lead-halide, the organic cation, water, and oxygen). A similar mechanism could be responsible for the increase in PLQE in MAPbBr$_3$ films, with the smaller overall healing as compared to I-based films resulting from the shorter Br-Pb bond length, subsequent lattice stabilization, and lower prevalence of defect states deep in the band gap[2,3,35,36].

When comparing the effect of solution stoichiometry on the properties of MAPbBr$_3$ perovskites and those of MAPbI$_3$, two key differences stand out. First, the microstructure of the former is highly sensitive to even minute changes in stoichiometry, while in the case of MAPbI$_3$ films nearly no variations in grain size or structure was observed. This should assist researchers in identifying MAPbBr$_3$ films made from solutions of slightly different stoichiometries and facilitate in increasing their reproducibility. The second stark difference is in the storage stability of the devices and since in both cases the solar cell architecture and extraction layers were kept the same, the differences are likely to originate from different active layer microstructures and densities of defect states. The microstructure of the perovskite active layer has been shown to influence its stability with degradation processes been shown to commence at grain boundaries in both $O_2$[9] and humid environments.[38] In the case of $MAPbI_3$, films of all stoichiometries exhibited the same microstructure and so it is predominantly the different densities of ionic defects that determined the storage stability of these devices. In the case of $MAPbBr_3$, both the microstructure and the density of defects vary, making assignment more complex. However, it is interesting to note that understoichiometric $MAPbBr_3$, which did not exhibit clear grain structure, are also the most stable, in agreement with the previously observed initiation of degradation at the grain boundaries. This suggests that for these samples, defect healing results in increased photovoltaic performance, while the smooth grain boundry-free microstructure delays the onset of degradation. However, given the high degree of overlapping phenomena between the microstructure, PV performance, PLQE, and their behavior over time, further research is needed to clearly elucidate the mechanisms at play in this healing behavior of the films. What our experiment clearly shows, however, is that the understoichiometric films display a high degree of improvement over time, indicating that the manipulation of defect states within the bulk film could be an effective strategy for increased film and device lifetime.

## 3. Conclusion

In conclusion, we examined the properties, performance, and shelf stability of $MAPbBr_3$ films and devices as a function of changing precursor solution stoichiometry. While the initial PV performance is constant over changing *y*, the films with a small bromide deficiency undergo a large increase in performance, with their PCE more than doubling in value from 3.5 % to nearly 8 %. Films with excess bromine also improve with prolonged shelf storage, but only by 1%. The maximum ELQE shows a similar trend. PLQE measurements under nitrogen and dry air measurements indicate that there is likely a trap healing mechanism at play, though the details of such a process remain to be closely examined. Though the overall trends are different than those of the MAPbI$_3$ films, both sets of results indicate the strong role that compositional variation plays in device longevity and film properties, suggesting that purposeful integration of defects into perovskite films may promote their eventual use.

## 4. Experimental Section

*Sample and Device Fabrication*: MAPbBr$_3$ precursor solutions were prepared following the method introduced in previous works[16,17]. In short, MABr was mixed with lead acetate trihydrate (PbAc$_2$•3H$_2$O) at an understoichiometric ratio and the stoichiometry was changed by adding specific amounts of MABr to the solution. A detailed explanation of this procedure can be found in Supplementary Note 1. Photovoltaic devices were fabricated using pre-patterned Glass/ITO substrates (PsiOTech Ltd.) that were cleaned by ultra-sonication in acetone and 2-propanol for 10 minutes each. The substrates were then blow-dried and O$_2$ plasma cleaned (100 W, 0.4 mbar) for 10 minutes. PEDOT:PSS (Heraeus) was spin-coated on the substrates at 4000 rpm for 45 s and annealed on a hot plate at 150 °C for 10 minutes in the ambient. The samples were then transferred into a dry-air glovebox in which the MAPbBr$_3$ active layers were deposited by spin-coating at 2000 rpm for 60 seconds. Immediately after spin-coating the samples were blow-dried with a dry air gun for 30 seconds and left to dry on an aluminum holder at room temperature for 5 minutes. Afterwards, the samples were annealed for 5 minutes at 80°C. Next, the samples were transferred into a N$_2$-filled glovebox, in which PCBM (20 mg/mL in chlorobenzene) electron transporting was spin-coated dynamically at 2000 rpm for 45 seconds, followed by a 10 min annealing at 80 °C. A BCP (0.5 mg/mL in 2-Propanol) hole-blocking layer was spin-coated at 4000 rpm for 25 seconds. The devices were completed by a thermal evaporation of a 80nm thick Ag electrode. Samples for spectroscopic and microscopic measurements were fabricated in an identical fashion to the active layer of the PV devices, but using glass substrates for UV-vis and PL and Glass/ITO/PEDOT:PSS for SEM, XPS and UPS.

*Device Characterization*: The devices were characterized as solar cells under simulated AM 1.5 sunlight at 100 mW cm$^{-2}$ irradiance (Abet Sun 3000 Class AAA solar simulator) with a Keithley 2450 source measure unit. The light intensity was calibrated with a Si reference cell (NIST traceable, VLSI) and corrected by measuring the spectral mismatch between the solar spectrum, the spectral response of the perovskite solar cell and the reference cell. The mismatch factor was calculated to be around 10%. When characterized as LEDs, the devices were measured inside an integrating sphere (Labsphere Inc.). The current−voltage characteristics were measured using a source-measure unit (Keithley 2450). At the same time the emitted light spectra were recorded using a scientific grade spectrometer (Ocean OpticsQE65Pro). The optical system (integrating sphere, spectrometer, and coupling optical fiber) were calibrated using a calibrated light source (Ocean Optics HL-2000-CAL).

*Photoemission Spectroscopy:* The MAPbBr$_3$ samples were transferred into an ultrahigh vacuum (UHV) chamber of the PES system (Thermo Scientific ESCALAB 250Xi) for measurements. The samples were exposed to air only for a short time span of approximately 30 seconds. All measurements were performed in the dark and several spots on each sample were measured in order to ensure enough statistics. Ultraviolet photoelectron spectroscopy measurements were carried out using a double-differentially pumped He discharge lamp (hν = 21.22 eV) with a pass energy of 2 eV and a bias at −10 V. XPS measurements were performed using an XR6 monochromated Al Kα source (hν = 1486.6 eV) and a pass energy of 20 eV.

*UV-Vis Spectroscopy:* Optical absorption spectra were measured with a Jasco UV-660 spectrophotometer in the range from 400 to 700 nm. The absorption of the substrate was subtracted as a baseline correction.

*Photoluminescence Spectroscopy:* PLQE measurements were carried out inside an integrating sphere (LabSphere) with excitation by a 405 nm CW laser (Coherent). The spectra were recorded using a QE65 Pro (Ocean Optics) spectrometer.

*Scanning Electron Microscopy:* SEM imaging was performed using a JSM-7610F FEG-SEM (Jeol). Samples were mounted on standard SEM holders using conductive Ag paste to avoid sample charging. The images were recorded using the secondary electron detector (LEI) at an acceleration voltage of 1.5 kV and a chamber pressure $<10^{-6}$ mbar.

**Acknowledgements**

The authors would like to kindly thank Prof. U. Bunz for providing access to film fabrication facilities and Prof. J. Zaumseil for access to SEM. We kindly thank the DFG (VA 991/2-1) for funding. This project has also received funding from the European Research Council (ERC) under the European Union's Horizon 2020 research and innovation programme (ERC Grant Agreement no. 714067, ENERGYMAPS).

**References**

[1]    NREL, "Best research-cell efficiencies," can be found under http://www.nrel.gov/ncpv/, **n.d.**

[2]    J. M. Ball, A. Petrozza, *Nat. Energy* **2016**, *1*, 16149.

[3]    A. Buin, P. Pietsch, J. Xu, O. Voznyy, A. H. Ip, R. Comin, E. H. Sargent, *Nano Lett.* **2014**, *14*, 6281.

[4]    B. Rivkin, P. Fassl, Q. Sun, A. D. Taylor, Z. Chen, Y. Vaynzof, *ACS Omega* **2018**, *3*, 10042.

[5]    S. N. Habisreutinger, T. Leijtens, G. E. Eperon, S. D. Stranks, R. J. Nicholas, H. J. Snaith, *Nano Lett.* **2014**, *14*, 5561.

[6]    S. K. Pathak, A. Abate, P. Ruckdeschel, B. Roose, K. C. Gödel, Y. Vaynzof, A. Santhala, S. I. Watanabe, D. J. Hollman, N. Noel, A. Sepe, U. Wiesner, R. Friend, H. J. Snaith, U. Steiner, *Adv. Funct. Mater.* **2014**, *24*, 6046.

[7]    C. Besleaga, L. E. Abramiuc, V. Stancu, A. G. Tomulescu, M. Sima, L. Trinca, N.

Plugaru, L. Pintilie, G. A. Nemnes, M. Iliescu, H. G. Svavarsson, A. Manolescu, I. Pintilie, *J. Phys. Chem. Lett.* **2016**, *7*, 5168.

[8] N. Phung, A. Abate, *Small* **2018**, *14*, 1802573.

[9] Q. Sun, P. Fassl, D. Becker-Koch, A. Bausch, B. Rivkin, S. Bai, P. E. Hopkinson, H. J. Snaith, Y. Vaynzof, *Adv. Energy Mater.* **2017**, *7*, 1700977.

[10] Q. Wang, B. Chen, Y. Liu, Y. Deng, Y. Bai, Q. Dong, J. Huang, *Energy Environ. Sci.* **2017**, *10*, 516.

[11] M. Saliba, T. Matsui, J.-Y. Seo, K. Domanski, J.-P. Correa-Baena, M. K. Nazeeruddin, S. M. Zakeeruddin, W. Tress, A. Abate, A. Hagfeldt, M. Grätzel, *Energy Environ. Sci.* **2016**, *9*, 1989.

[12] F. Bella, G. Griffini, J. P. Correa-Baena, G. Saracco, M. Grätzel, A. Hagfeldt, S. Turri, C. Gerbaldi, *Science (80-. ).* **2016**, *354*, 203.

[13] Z. Chen, X. Zheng, F. Yao, J. Ma, C. Tao, G. Fang, *J. Mater. Chem. A* **2018**, *6*, 17625.

[14] T. A. Berhe, W. N. Su, C. H. Chen, C. J. Pan, J. H. Cheng, H. M. Chen, M. C. Tsai, L. Y. Chen, A. A. Dubale, B. J. Hwang, *Energy Environ. Sci.* **2016**, *9*, 323.

[15] Q. Sun, P. Fassl, Y. Vaynzof, *ACS Appl. Energy Mater.* **2018**, *1*, 2410.

[16] P. Fassl, V. Lami, A. Bausch, Z. Wang, M. T. Klug, H. J. Snaith, Y. Vaynzof, *Energy Environ. Sci.* **2018**, *11*, 3380.

[17] P. Fassl, Y. Zakharko, L. M. Falk, K. P. Goetz, F. Paulus, A. D. Taylor, J. Zaumseil, Y. Vaynzof, *J. Mater. Chem. C* **2019**, *7*, 5285.

[18] J. H. Heo, S. H. Im, *Adv. Mater.* **2016**, *28*, 5121.

[19] T. Leijtens, K. A. Bush, R. Prasanna, M. D. McGehee, *Nat. Energy* **2018**, *3*, 828.

[20] M. Jaysankar, B. A. L. Raul, J. Bastos, C. Burgess, C. Weijtens, M. Creatore, T. Aernouts, Y. Kuang, R. Gehlhaar, A. Hadipour, J. Poortmans, *ACS Energy Lett.* **2019**, *4*, 259.

[21] C. Aranda, A. Guerrero, J. Bisquert, *ACS Energy Lett.* **2019**, *4*, 741.

[22]  H. Cho, S.-H. Jeong, M.-H. Park, Y.-H. Kim, C. Wolf, C.-L. Lee, J. H. Heo, A. Sadhanala, N. Myoung, S. Yoo, S. H. Im, R. H. Friend, T.-W. Lee, *Science (80-. ).* **2015**, *350*, 1222.

[23]  S. J. Kim, J. Byun, T. Jeon, H. M. Jin, H. R. Hong, S. O. Kim, *ACS Appl. Mater. Interfaces* **2018**, *10*, 2490.

[24]  V. Prakasam, F. Di Giacomo, R. Abbel, D. Tordera, M. Sessolo, G. Gelinck, H. J. Bolink, *ACS Appl. Mater. Interfaces* **2018**, *10*, 41586.

[25]  J. C. Yu, D. Bin Kim, E. D. Jung, B. R. Lee, M. H. Song, *Nanoscale* **2016**, *8*, 7036.

[26]  R. K. Misra, S. Aharon, B. Li, D. Mogilyansky, I. Visoly-Fisher, L. Etgar, E. A. Katz, *J. Phys. Chem. Lett.* **2015**, *6*, 326.

[27]  R. Ruess, F. Benfer, F. Böcher, M. Stumpp, D. Schlettwein, *ChemPhysChem* **2016**, *17*, 1505.

[28]  W. Zhang, M. Saliba, D. T. Moore, S. K. Pathak, M. T. Hörantner, T. Stergiopoulos, S. D. Stranks, G. E. Eperon, J. A. Alexander-Webber, A. Abate, A. Sadhanala, S. Yao, Y. Chen, R. H. Friend, L. A. Estroff, U. Wiesner, H. J. Snaith, *Nat. Commun.* **2015**, *6*, 6142.

[29]  W. Zhang, S. Pathak, N. Sakai, T. Stergiopoulos, P. K. Nayak, N. K. Noel, A. A. Haghighirad, V. M. Burlakov, D. W. DeQuilettes, A. Sadhanala, W. Li, L. Wang, D. S. Ginger, R. H. Friend, H. J. Snaith, *Nat. Commun.* **2015**, *6*, 10030.

[30]  D. Forgács, M. Sessolo, H. J. Bolink, *J. Mater. Chem. A* **2015**, *3*, 14121.

[31]  W. Nie, H. Tsai, R. Asadpour, J.-C. Blancon, A. J. Neukirch, G. Gupta, J. J. Crochet, M. Chhowalla, S. Tretiak, M. A. Alam, H.-L. Wang, A. D. Mohite, *Science (80-. ).* **2015**, *347*, 522.

[32]  A. Buin, R. Comin, J. Xu, A. H. Ip, E. H. Sargent, *Chem. Mater.* **2015**, *27*, 4405.

[33]  N. Aristidou, C. Eames, I. Sanchez-Molina, X. Bu, J. Kosco, M. S. Islam, S. A. Haque, *Nat. Commun.* **2017**, *8*, 15218.

[34]  N. Aristidou, I. Sanchez-Molina, T. Chotchuangchutchaval, M. Brown, L. Martinez, T. Rath, S. A. Haque, *Angew. Chemie - Int. Ed.* **2015**, *54*, 8208.

[35]  A. Jaffe, Y. Lin, C. M. Beavers, J. Voss, W. L. Mao, H. I. Karunadasa, *ACS Cent. Sci.* **2016**, *2*, 201.

[36]  T. Shi, W. J. Yin, F. Hong, K. Zhu, Y. Yan, *Appl. Phys. Lett.* **2015**, *106*, DOI 10.1063/1.4914544.

[37]   Q. Wang, B. Chen, Y. Liu, Y. Deng, Y. Bai, Q. Dong and J. Huang, *Energy Environ. Sci.* **2017**, 10, 516–522.

In this work, the effect of variations in the precursor stoichiometry on the properties, performance and stability of MAPbBr$_3$ photovoltaic devices is examined. It is found that while the initial device performance is similar, devices with understoichiometric composition are significantly improved upon storage in air without illumination, which is associated with defect healing.

**Keyword** Lead bromide perovskites, photovoltaic devices, stability, reproducibility, stoichiometry

*Lukas M. Falk, Katelyn P. Goetz, Vincent Lami, Qingzhi An, Paul Fassl, Jonas Herkel, Fabian Thome, Alexander D. Taylor, Fabian Paulus and Yana Vaynzof\**

**Effect of Precursor Stoichiometry on the Performance and Stability of MAPbBr$_3$ Photovoltaic Devices**

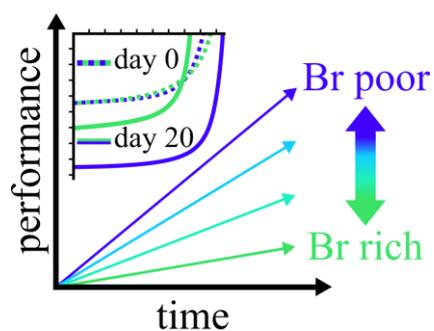